\newcommand{\xuv}{{\sc xuv}}
\newcommand{\ir}{{\sc ir}}

\newcommand{\iap}{{\sc iap}}

\newcommand{\ewp}{{\sc ewp}}

\newcommand{\ddaa}{{\sc ddaa}}

\newcommand{\tdse}{{\sc tdse}}

\newcommand{\ata}{{\sc ata}}

\documentclass[twocolumn,showpacs,preprintnumbers,amsmath,amssymb]{revtex4}
\usepackage{graphicx}
\usepackage{dcolumn}
\usepackage{bm}
\usepackage{color}

\begin{document}
\bibliographystyle{revtex}
\title{
Quantum interference in attosecond transient absorption\\
of laser-dressed helium atoms
}

\author{Shaohao Chen}
\author{Mengxi Wu}
\author{Mette B. Gaarde}
\author{Kenneth J. Schafer}

\affiliation{
Department of Physics and Astronomy, Louisiana State University, Baton Rouge 70803, USA
}

\begin{abstract}
We calculate the transient absorption of an isolated attosecond pulse
by helium atoms subject to a delayed  infrared (\ir) laser pulse. 
With the central frequency of the broad attosecond  spectrum  near the  ionization threshold, the absorption spectrum  is strongly modulated at the sub-\ir-cycle level. 
Given that the absorption spectrum results from a time-integrated measurement, we  investigate the extent to which the delay-dependence of the absorption  yields information about the attosecond dynamics of the atom-field energy exchange. 
We find two configurations in which this is possible. The first involves multi photon transitions between bound states that result in  interference between different excitation pathways. The other involves the modification of the bound state absorption lines by the \ir\ field, which we find can result in a sub-cycle time dependence only when ionization limits the duration of the strong field interaction.

\end{abstract}

\date{December 29, 2012}
\maketitle
Attosecond transient absorption (\ata) studies provide a way to push our understanding of the energy transfer between electromagnetic fields and matter to the sub-femtosecond time scale \cite{Gou10739,PhysRevLett.105.143002}. They are an all-optical attosecond metrology that complements methods based on the measurement of charged particles, such as  attosecond streaking \cite{PhysRevLett.88.173903,RevModPhys.81.163} and electron interferometry \cite{Rem06323,PhysRevLett.105.053001}. Like those methods, high time-resolution is gained  in an \ata\ spectrometer by using attosecond extreme ultraviolet (\xuv) pulses that are synchronized to the field oscillations of an infrared (\ir) laser pulse. 
The first \ata\ experiments used an attosecond pulse to probe a valence electron wave packet (\ewp) created by ionizing an atom with a strong \ir\ laser pulse \cite{Gou10739, Wirth14102011}. Recent \ata\ experiments  have used attosecond pulses as a {\it pump} that creates an \ewp\ which is then probed by a moderately strong  \ir\ field.  In this \xuv-pump/\ir-probe configuration the frequency-resolved transient absorption signal varies as a function of the \ir\ intensity, duration, and the sub-cycle timing between the two fields \cite{PhysRevLett.106.123601, ISI:000305410500014, PhysRevLett.109.073601}.  These experiments raise the possibility of studying time-dependent absorption down to the attosecond time scale.

In this theoretical study, similar to recent experiments \cite{PhysRevLett.109.073601,PhysRevA.86.063408}, we consider helium atoms that are excited by an isolated attosecond pulse (\iap) with a central frequency near the ionization threshold, together with a delayed few-cycle \ir\  pulse.
We elucidate the key 
features  of the resulting delay-dependent attosecond absorption (\ddaa) spectra. These features derive from 
 the fact that the \iap\  ``starts the clock'' by exciting the system at a well-defined time, and that it makes an \ewp\  that is essentially independent of the pump-probe delay. The \ddaa\ measurement is  a spectrogram that records the interference between different excitation pathways which lead to the same absorption or emission processes. 
Because the attosecond pulse is locked to the \ir\ field oscillations, it is suggestive that many of the features in the spectrum are modulated at half the \ir\ laser period ($T_{\rm IR}$), about 1.3 fs.  
However, the absorption spectrum results from the time-integrated response of everything that happens after the initial \xuv\ excitation, which complicates the analysis.  Our main concern in this paper is the connection between these oscillations and the real-time attosecond dynamics.

We discuss two distinct manifestations of attosecond dynamics in the \ddaa\ spectrum. First, near the ionization threshold the delay-dependent absorption exhibits fully modulated interference fringes at  half the \ir\ period. We show these are due to the interference between two pathways, separated in time, which give rise to the same dipole response. This ``which way'' quantum path interference  \cite{Paul961111} can be used to time-resolve two photon transitions between excited states, and shows how the absorption at \xuv\ frequencies is altered on a sub-cycle time scale.
We   find  $T_{\rm IR}/2$ oscillations with a similar origin in absorption features associated with non-linear \xuv+\ir\ processes, which appear as light-induced structures in the  \ata\ spectrum. These fringes illustrate how energy can be exchanged between excitation modes (dressed states) that exist only when the \xuv\ and \ir\ fields overlap in time.
The second feature results from the \ir\ driven sub-cycle AC Stark shift of  the lower-lying  $2p$ and $3p$ resonances \cite{PhysRevLett.109.073601}. 
The time-dependent Stark shift  leads to a dispersive line-shape, which results from interference between the time-dependent dipole induced before, during, and after the \ir\ pulse. We show that, due to the integrated nature of the absorption spectrum, a measurement of the instantaneous energy shift is only possible when ionization limits the interaction time.  

The essential features of \ddaa\ spectra can be understood at the single atom level. 
The energy lost or gained by a light field in the interaction with an atom 
can be described by a frequency-dependent response function 
$\tilde{S}(\omega,t_d)=2\, {\rm Im}[\tilde{d}(\omega) \tilde{\cal E}^* (\omega)]$,
where $t_d$ is the pump-probe delay (we use atomic units unless otherwise indicated).
$\tilde{d}(\omega)$ and $\tilde{\cal E}(\omega)$ are the Fourier transforms  
of the time-dependent dipole moment $d(t)$ and the total driving field $E(t)$.  The dipole moment is obtained by solving the time-dependent Schr\"{o}dinger equation (\tdse) 
in the single-active-electron approximation 
\cite{Gaa11013419}.

Fig.~1 shows a typical response function vs pump-probe delay. 
The 25 eV, 330 as  \iap\ pump 
pulse has a bandwidth of 5.5 eV, which means that it overlaps  all the singly excited and low-energy continuum states of the He atom. 
The FWHM of the 800 nm probe pulse is 11 fs, corresponding to 4 optical cycles  ($T_{\rm IR} = 2.7$~fs). The \ir\ field is of moderate intensity, so that it can not itself excite the atom in its ground state. The pump and probe fields have parallel linear polarizations. We include a dipole dephasing time $T_2= 65$ fs in the calculations by smoothly windowing the dipole moment. We have verified that using a longer $T_2$ does not change any of our conclusions.  
Positive (negative) values of $\tilde{S}(\omega,t_d)$ mean that a dilute gas will absorb (emit) energy at frequency $\omega$
as the dipole-driven source term in the wave equation will be out of (in) phase with $\tilde{\cal E}(\omega)$ \cite{Gaa11013419,PhysRevA.85.013415}.
\begin{figure}[t]
\centering
\includegraphics[width=0.50\textwidth, trim=10mm 0mm 0mm 0mm, clip]{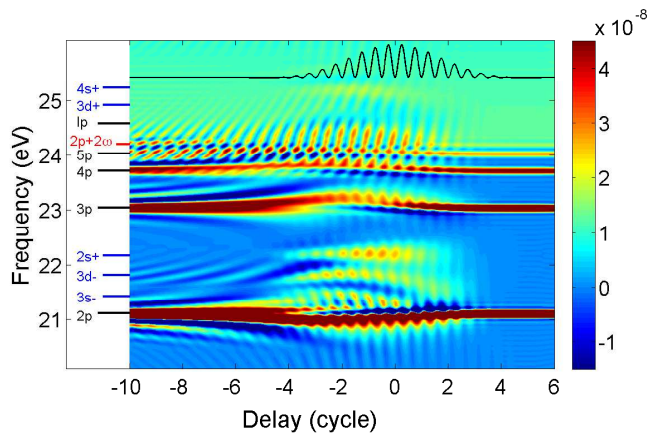}
\caption{\label{fig1} \small
Single atom response function $\tilde{S}(\omega,t_d)$  in helium where $t_d$ is the time delay in \ir\ cycles
between the \ir\ laser pulse 
($800$ nm, $3 \times 10^{12}$ W/cm$^2$, 4 cycles, $\cos^2$ envelope, sine-like carrier envelope)
and the attosecond pulse (330 as, centered at 25 eV).
The \ir\ intensity oscillations are shown in black in the top panel.
}
\end{figure}

We begin by briefly describing the main features in Fig.~1. 
For large positive delays, $t_d  \ge 4\,T_{\rm IR}$, the \iap\ arrives  after the end of \ir\ pulse and the absorption spectrum exhibits only one-\xuv-photon transitions from the ground state to the n$p$ states. When the two pulses overlap, $-4\,T_{\rm IR} < t_d < 4\,T_{\rm IR}$,  these absorption features are strongly modified.
In this same delay range, light-induced structures (LISs) appear, for example, between the $2p$ and $3p$ resonant lines. These are associated with  two-photon ({\xuv} $\!\pm\!$ {\ir}) processes that transfer population from the ground state to non-dipole coupled (``dark'') $s$ and $d$ states, and have recently been observed experimentally  \cite{PhysRevA.86.063408}. 
Finally, when the \iap\ arrives before the \ir\  pulse, $t_d \le-4\,T_{\rm IR}$, the dipole established by the \xuv\ pulse undergoes free decay until the \ir\ pulse arrives and strongly modified the dipole, which induces sidebands on the main resonance features.

\begin{figure}[t]
\centering
\includegraphics[width=0.48\textwidth, trim=15mm 2mm 0mm 0mm, clip]{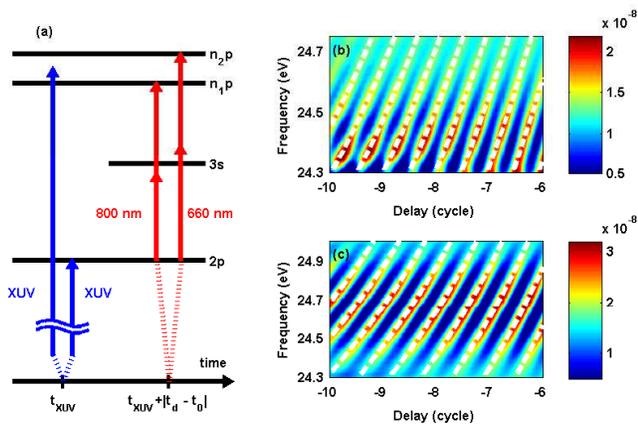}
\caption{\label{phase} \small
(a) Diagram of direct (one-photon) and indirect (three-photon) pathways 
to the same final $n_1 p$ (or $n_2 p$) states.
Comparison of absorption maxima and the constructive interference condition (white dashed curves) 
for a narrow range of frequencies and \ir\ wavelengths of 
(b) 800 nm (same as Fig.~1) and (c) 660 nm.
}
\end{figure}

Many of the features in $\tilde{S}(\omega, t_d)$ show a modulation at one half the laser period.  Of particular interest are fringes that are present both when the pulses overlap and when the \iap\ arrives before the \ir.
An example is seen in Fig.~1 between 24.2 and 24.8 eV near the label ``$2p+2\omega$'', also shown in more detail in Fig.~\ref{phase}(b). These fringes 
 are caused by  quantum interference 
between two distinct pathways for establishing the same coherence between the ground state 
and a group of n$p$ states (${\rm n}>5$), near threshold. The process is  diagrammed in Fig.~\ref{phase}(a). 
The direct pathway is an \xuv-driven 1st-order process 
that populates all of the n$p$ states simultaneously in a manner that is independent of the delay. 
The indirect  pathway is a 3rd-order process in which amplitude in the 2$p$ states is transferred to an n$p$ state by a two-\ir-photon process. 
The role of the indirect pathway via the $2p$ state is confirmed by a test calculation which dynamically eliminates the $2p$ state during the time-propagation of the \tdse, in which these fringes disappear.
The two interfering processes, since they are driven by the \xuv\ field and the \ir\ field respectively, 
happen at different times depending on the value of $t_d$. 
The resulting interference fringes are analogous to those  observed in photoelectron spectra  \cite{PhysRevLett.105.053001}, however, 
in \ddaa\ the multiphoton process need not result in ionization. 

Having identified the interfering pathways  it is straightforward to write down
the constructive interference condition as a function of delay:
\begin{eqnarray}
\label{condition} 
(\omega_{np}-\omega_{2p}) |t_d-t_0| + \Delta \phi = 2\pi k,
\end{eqnarray}
where $\omega_{np}$ and $\omega_{2p}$ are the energies of the $np$  and $2p$ states respectively,  
 and $k$ is a positive integer \cite{PhysRevLett.105.053001}.
Two additional parameters enter:  $\Delta \phi$ is a phase due to the two-\ir-photon transition and
 $t_0$ is the time when the two photon transition probability peaks. It is reasonable to choose $t_0=-T_{\rm IR}/4$ due to the sine-like \ir\ carrier wave used in the calculation. 
This formula correctly predicts a decreasing slope for the fringes as $|t_d-t_0|$ increases,
corresponding to increasing $k$ in Eq.~(\ref{condition}).
The fit to the interference fringes using $\Delta \phi=0$, one set of integers $k$ that start with $k=1$ at $t_0$,  
and no other free parameters, is excellent for $t_d < 0$, as shown in Fig. \ref{phase}(b). 
For $t_d > 0$ the biggest contribution to the indirect pathway comes from 
whatever \ir\ field maximum follows the \xuv\ pulse (never more than one half cycle away). The slope of the fringes should therefore saturate for $t_d > 0$, which is indeed what we observe in Fig.~1. We can thus conclude that the strength of the two photon transition peaks at the local intensity maxima of the \ir\ field. This supports a time-dependent picture of multi-photon absorption below threshold as a process that follows the sub-cycle oscillations of the \ir\ electric field.

One application of this 1st vs.\ 3rd order interference process is illustrated in Fig. \ref{phase}(a). Realizing that  the indirect pathway is enhanced due to 
a near one-\ir-photon resonance with the intermediate $3s$ state, it
seems possible that we can use this  to determine $\Delta\phi$ in Eq. (\ref{condition}).
For an 800 nm  pulse we can safely assume that $\Delta \phi=0$ 
because the intermediate $3s$ state lies above the energy reached 
by absorbing one \ir\ photon from the $2p$ state. 
In Fig. \ref{phase}(c) we have repeated the calculation in Fig.~1 using a wavelength of 660 nm.  
Since the absorption of one photon now reaches an energy above the $3s$ state, 
we expect this phase should shift by $\pi$. 
In Fig. \ref{phase}(c) we use $\Delta\phi =\pi$ for the same $t_0$ and 
achieve an excellent fit to the calculation. 
The interference fringes can therefore be used to measure the phase of an intermediate resonance 
if the pulse wavelength can be scanned~\cite{PhysRevLett.104.103003}. Another, more speculative application, is 
as a probe of decoherence. A short decoherence time 
would reduce the fringe contrast for large negative delays, from which the time scale for decoherence could be inferred directly.  

We can apply a similar analysis  to the LISs as well.
In the overlap region ($-4\,T_{\rm IR} < t_d < 4\,T_{\rm IR}$) 
each LIS is associated with an $s$ or $d$ bound state which is coupled to the ground state by one \xuv\ photon and the absorption or emission of an additional \ir\ photon. We use a plus or minus label for LIS  that are one photon above or below the final $s$ or $d$ state. 
LISs associated with coupling of the ground state to the  $3s$, $3d$, and $2s$ states can be seen near 21.3 eV
($3s_{-}$), 21.7 eV ($3d_{-}$),  and 22.1 eV ($2s_{+}$), respectively.
Similar LISs are visible above threshold ($3s_+$, $3d_+$ and $4s_+$).
The half cycle modulations in the LISs come from the 2$\omega_{\rm IR}$ coupling of plus and minus LISs, which have an atomic state as an intermediate resonance. For example, the $3d_-$ is two-photon resonantly coupled to the $3d_{+}$ (visible near 25 eV around $t_d=-1\,T_{IR}$), yielding oscillations that are 
are exactly out of phase with each other.
These couplings can also be confirmed by examining the tilt of the LIS interference fringes.
Eq.~(1) predicts that the fringes of the $3d_-$ have a negative slope with respect to delay, which is indeed the case in Fig. 1.
However, the fringes in $2s_+$ LIS have a positive slope, characteristic of being above the final $2s$ state.
Fits to Eq.~(1) using the LIS energies  show that the two photon transitions between LISs also follow the \ir\ electric field oscillation.

A time-dependent perspective on these oscillations is provided by the observation that the coupling between the one-\xuv-photon resonant n$p$ state and  nearby dark states is in a regime where the rotating wave approximation breaks down, and that the short duration of the \iap\ yields a well-defined phase of the \ir\ field at excitation. 
 In such situations it has been demonstrated that the final state populations are sensitive to the  phase of the laser at the time of excitation \cite{PhysRevA.69.032308}.
 Here we have shown that the \ddaa\ spectrum shows the same sensitivity. 

Another source of fast dynamics in the transient absorption spectrum is the sub-cycle AC Stark shift of the bound state energies  \cite{PhysRevLett.109.073601}.
Figs. 3(a) and (b) show the evolution of the $2p$ and $3p$ line shapes with delay, which include both emission and absorption. Qualitatively, this results from the redistribution of \xuv\ energy across  the resonances due to the perturbation of the dipole by the \ir\ field. Similar dispersive line shapes have been discussed previously in connection with the control of exciton polarizations of a quantum dot 
using picosecond laser pulses \cite{PhysRevLett.92.157401,PhysRevLett.89.057401}.

In a time-domain picture, the dispersive shape of the n$p$ absorption lines is caused by  multiple contributions to the dipole moment: (i) a perturbed response during 
the \ir\ pulse,
 and (ii) free decay of the coherence after the \ir\ pulse ends \cite{PhysRevB.38.7607,PhysRevB.48.4695}. The relative importance of these terms is determined by 
the \ir\ field intensity and delay, as well as $T_2$. If the state amplitude is unchanged by the \ir\ field then the main contribution 
to the dipole moment comes from the free decay which, however, acquires a phase shift, $\theta_{\rm S }$, equal to the integrated optical Stark shift during the time when the \ir\ pulse acts.
In this approximation the response is given by
\begin{equation}
\tilde{S}(\omega,t_d)\approx {\cal L}(\omega,T_2)\left[ \cos(\theta_{\rm S}) +(\Delta \cdot T_2)\sin(\theta_{\rm S})\right]
\end{equation}
where $\Delta$ is the difference $\omega-\omega_{{\rm n}p}$ and ${\cal L}(\omega,T_2)$ is the Lorentzian line shape in the absence of the \ir\ field.
For positive delays $1 \leq t_d < 4T_{\rm IR}$ we find that the line shape is well described by this simple form. We see that the shift of the absorption is in the {\it direction} of the Stark shift, but the magnitude of the shift is proportional to $\theta_{\rm S}/T_2$ and is therefore constrained by the natural linewidth. Thus, for these delays,  the 2$p$ and 3$p$ lines have Stark shifts of opposite sign, as shown in Fig.~3(c), but the magnitude of these shifts does not generally equal the instantaneous Stark shift.

\begin{figure}
\centering
\includegraphics[width=0.5\textwidth, trim=35mm 8mm 25mm 8mm, clip]{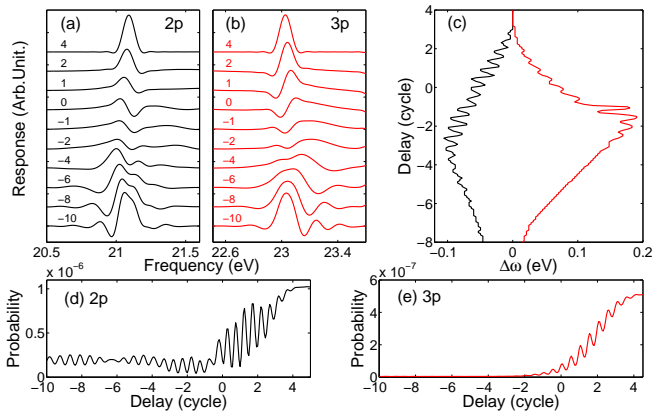}
\caption{\small
Response as a function of frequency at around 2p (a) and 3p (b) resonance (lineouts of Fig. 
1 at different delays). Energy shifts retrieved from the maximum response (c) for 2p (black) and 3p 
(red) states. Half-cycle modulation of absorption strength shown in integration of response function 
(black and red), as well as in final populations at the end of the pulses (green and blue) for 2p (d) and 
3p (e) sates.
}
\label{emission}
\end{figure}

At delays near zero ($-4\,T_{\rm IR} \leq t_d<1\,T_{\rm IR}$)  the lineshape becomes much wider than 1/$T_2$ and more complex than the simple dispersive shape discussed above. Here depopulation of the excited states by the \ir\ field plays a larger role 
(see Fig.\ 3(d-e)), and the response is dominated by the perturbed dipole decay. 
 For the 3$p$ state the loss is predominantly by one photon ionization to the continuum and the dipole response exists for only a few laser cycles after it is excited. This rapid ionization-induced dephasing allows us to probe the real-time sub-cycle dynamics using a delay-dependent method: during this time (near -2\,$T_{\rm IR}$ in Fig. 3(c)) we observe a large transient shift  of the 3$p$ feature of about 0.19 eV, 
 comparable to the ponderomotive energy of  0.18 eV, and strong half cycle oscillations in the peak position, showing the rapid variation in the energy of the resonance as it follows the \ir\ field oscillations. 
These observations are in line with other investigations which show that the Stark shift is difficult to observe optically in the absence of rapid dephasing \cite{Wirth14102011}. This means that observing fast dynamics via ionization-induced dephasing requires careful control of the \ir\ intensity. In the continuum, however, all states have a very short dephasing time, so we  always observe an overall ponderomotive shift of the absorption, though the response is very weak.

The behavior of the peak of the $2p$ lineshape is more complex than the 3$p$, in part because the 2$p$ is never fully depopulated during the \ir\ pulse (Fig. 3 (d)), and in part because of the strong coherent coupling between the 2$p$ state and the nearby dark states. The 2$p$ lineshape therefore always results from both perturbed and free decay. The sub-cycle oscillations of the $2p$ position are most likely due to the which-way interference with the $np$ states discussed in the context of Fig.~2, judging from the negative and changing slope of the fringes between $-3\,T_{\rm IR}$ and  $3\,T_{\rm IR}$ (as seen in Fig.~1). As such they do {\it not} contain information about the sub-cycle AC Stark shift.

For delays $t_d < -4T_{\rm IR}$ the \iap\ and \ir\ act  separately: the dipole established by the \iap\ oscillates freely until the \ir\ pulse strongly perturbs it, greatly altering  the dipole amplitude and phase in a few \ir\ cycles. 
This perturbed free polarization decay has been  observed previously at optical frequencies \cite{Cruz88261,PhysRevB.38.7607}. It yields sidebands, seen clearly above the 3$p$ and below the 2$p$ lines, with a characteristic hyperbolic shape that depends on the delay as $1/t_d$ and shows no attosecond time scale dynamics. 

To conclude, we have  identified two configurations where attosecond dynamics can be observed in attosecond transient absorption spectra: which-way interference and the sub-cycle AC Stark shift.  Since experimental observations always result from propagation in a macroscopic gas, 
we also solve the Maxwell wave equation ({\sc mwe})
for the time propagation of light fields through the atomic helium gas medium,
in which the polarization and ionization source terms 
are obtained by solving the single-atom \tdse\ \cite{Gaa11013419}.
The absorption/emission probability after propagation of
a dilute gas is almost identical to the single atom result in Fig.~1, exhibiting both the half-cycle modulations, the LISs, and the emission features discussed above. 
This indicates that our \ddaa\ predictions can be observed using current attosecond technology, 
especially as it regards half-\ir-cycle oscillations. 

We have demonstrated that interference fringes in \ddaa\ spectra result from the coherent addition of two quantum paths that lead to the same dipole excitation. They reveal the time delay between the initial excitation and a later, \ir-field driven multiphoton transition. This was found to be true both  for transitions between bound states as well as  between  excitation modes (the LISs) that are observable only when the pulses overlap. We expect therefore that they will be a general feature in more complex systems  \cite{arXiv:1205.0519v1}, and could be observed 
between resonant states embedded in a continuum, as long as the lifetimes are longer that the \ir\ period. 
Using this argument in reverse, the interference fringes visible in the \ddaa\ spectrum could be used as a precise timing device, 
a probe of decoherence, or a phase meter when an intermediate resonance is involved.

\section*{Acknowledgements}
This work was supported by the National Science Foundation under Grant No. PHY-0701372 and No. PHY-1019071. High-performance computational resources were provided by Louisiana State University (www.hpc.lsu.edu) and Louisiana Optical Network Initiative (www.loni.org).


\end{document}